\documentclass[twocolumn,superscriptaddress,showpacs,preprintnumbers,amsmath,amssymb]{revtex4}

\usepackage{latexsym}
\usepackage{graphicx}
\usepackage{dcolumn}

\newcommand{\bra}[1]{\langle #1 |}

\newcommand{\ket}[1]{| #1 \rangle}

\def\etal{\textit{et al.{}}}

\newcommand{\im}{\dot{\iota}\,}
\begin{document}

\title{Generation of two-dimensional cluster states by using high-finesse
       bimodal cavities}

\author{D. Gon\c{t}a}
\email{gonta@physi.uni-heidelberg.de}
\affiliation{Max-Planck-Institut f\"{u}r Kernphysik, P.O.~Box
             103980, D-69029 Heidelberg, Germany}

\author{T. Radtke}
\email{tradtke@physik.uni-kassel.de} \affiliation{Institut f\"{u}r
Physik, Universit\"{a}t Kassel,
             Heinrich-Plett-Str. 40, D-34132 Kassel, Germany}

\author{S. Fritzsche}
\email{s.fritzsche@gsi.de} \affiliation{Department of Physical
Sciences, P.O.~Box 3000,
             Fin-90014 University of Oulu, Finland}%
\affiliation{GSI Helmholtzzentrum f\"{u}r Schwerionenforschung,
             D-64291 Darmstadt, Germany}%

\date{\today}

\begin{abstract}
We propose two novel schemes to generate the two-dimensional $2
\times N$ and $3 \times N$ cluster states by using a chain of
(two-level) Rydberg atoms in the framework of cavity QED. These
schemes work in a completely deterministic way and are based on the
resonant interaction of the atoms in a chain with a \textit{bimodal}
cavity that supports two independent modes of the photon field. We
demonstrate that a $2 \times N$ cluster state can be generated
efficiently with only a single bimodal cavity, while two such
cavities are needed to produce a $3\times N$ cluster state. It is
shown, moreover, how these schemes can be extended towards the
generation of $M \times N$ two-dimensional cluster states.
\end{abstract}


\maketitle

\section{Introduction}

Entanglement is known today as a key feature and resource of quantum
mechanics. It has been found important not only for studying the
nonlocal and nonclassical behavior of quantum particles but also for
a wide range of applications in quantum engineering and quantum
information theory \cite{nc}, such as super-dense coding \cite{bew},
quantum cryptography \cite{eke}, or the quantum search algorithm
\cite{gro}. Apart from the Bell states as prototypes of two-partite
entangled states, various attempts have been made recently in order
to produce and control entanglement also for more complex systems.
Owing to the fragile nature of most of these states, however, the
manipulation of larger quantum systems still remains a great
challenge for experiment and only a few proof-of-principle
implementations have been realized so far that generate entangled
states with more than two parties in a well-controlled manner.

Recently, Briegel and Raussendorf \cite{BR1} have introduced a novel
type of multi-partite entangled states. These (so-called)
\textit{cluster} states are known to exhibit a rather high
persistency and robustness of their entanglement with regard to
decoherence effects \cite{prl92a}. Apart from the fundamental
interest in these states \cite{pra71} and their use in quantum
communication protocols \cite{pla364}, the cluster states also form
the key ingredient for one-way quantum computations \cite{BR2}. In
general, a cluster state can be constructed from an array of
uncorrelated qubits by carrying out the following two steps: (i) the
preparation of each qubit in the superposition $\ket{+} \equiv
\left(\ket{0} + \ket{1} \right) /\sqrt{2}$, where $\ket{0}$ and
$\ket{1}$ refer to the distinguishable basis states of some given
two-level system, such as the spin projection of a spin-1/2 particle
or the polarization of light, and (ii) the (subsequent) application
of the controlled-z operation \cite{nc}
\begin{equation}\label{cz-def}
\ket{i}\ket{j} \:\longrightarrow\: (-1)^{ij} \ket{i}\ket{j}; \qquad
i,j = 0,1
\end{equation}
between--some or all--pairs of neighboring qubits in order to
entangle them with each other.

Since the original paper by Briegel and Raussendorf, the generation
of cluster states has attracted much attention and has become a
research topic by itself. Using a linear-optical set-up, for
example, a proof-of-principle implementation of a four-qubit cluster
state has first been reported by Walther and coworkers
\cite{nat434}, and was utilized also by Tokunaga \etal{}
\cite{tok08} in order to demonstrate basic operations for the
one-way quantum computing. In the framework of cavity QED, in which
neutral atoms are coupled to a high-finesse microwave or optical
cavity, different schemes have been suggested during recent years to
generate linear cluster states \cite{pra72, pra73, pra75, pra77a,
mps}. In contrast to the linear cluster states, the two-dimensional
cluster states would enable one to perform also multi-qubit gate
operations (e.g.~quantum gates that act on two or more qubits
simultaneously) in one-way computations \cite{BR2} and, therefore,
may result in a viable alternative to the conventional (circuit)
computations in which sequences of unitary gates need to be carried
out. Up to the present, nevertheless, only a minor progress has been
achieved in Ref.~\cite{oc281} with regard to schemes that generate
two-dimensional cluster states within cavity QED. In this reference,
the two-dimensional cluster state is obtained by combining two (or
more) linear cluster states, which however, requires demanding
experimental set-up and imposes certain restrictions on the geometry
of the output cluster state.

In this paper, we suggest two practical schemes for the generation
of two-dimensional $2 \times N$ and $3 \times N$ cluster states that
are feasible for modern cavity QED experiments. These schemes work
in a completely deterministic way and are based on the resonant
interaction of a chain of Rydberg atoms with (high-finesse) bimodal
cavities which, in contrast to a single-mode cavity, support two
independent modes of the light field. While only one of these
cavities is required for the generation of the $2 \times N$ cluster
states, two (and more) cavities are needed to construct cluster
states of larger size. Below, we describe the individual steps in
the interaction of the Rydberg atoms with the cavity modes that are
required to perform the suggested scheme. Here, we shall introduce
also a graphical language in order to display all these steps in
terms of quantum circuits and temporal sequences of the interactions
that each of atoms undergoes. After all the atom-cavity interactions
have been completed, the cluster state is encoded in the chain of
atoms that has passed through one (in the first scheme for the $2
\times N$ cluster) or through two subsequent cavities (in the second
scheme for the $3 \times N$ cluster). In addition, we also show how
the suggested procedure can be extended to construct two-dimensional
cluster states of arbitrary size, once a sufficiently large chain of
atoms and an array of cavities are provided. We briefly discuss the
implementation of one-way quantum computations within the given
set-up in order to demonstrate that our approach is well suited for
present-day experiments using bimodal microwave cavities. In
addition, we briefly point to and discuss the main limitations that
may arise experimentally in the generation of larger cluster states
by using microwave cavities similar to those as utilized in the
Laboratoire Kastler Brossel (ENS) in Paris.

The paper is organized as follows. In the next Section, we start
with a brief reminder on the (resonant) interaction of one two-level
Rydberg atom with a bimodal cavity. In Section II.A, we then recall
the steps that are necessary in order to generate the linear $1
\times N$ cluster state, and which sets the stage also to discuss
the generation of two-dimensional cluster states. In Section II.B,
we present the scheme for the $2 \times N$ cluster state and, in
Section II.C, for the $3 \times N$ state as well as for
two-dimensional clusters of larger size. Finally, a summary and
outlook are given in Section IV.

\begin{figure}
\begin{center}
\includegraphics[width=0.46\textwidth]{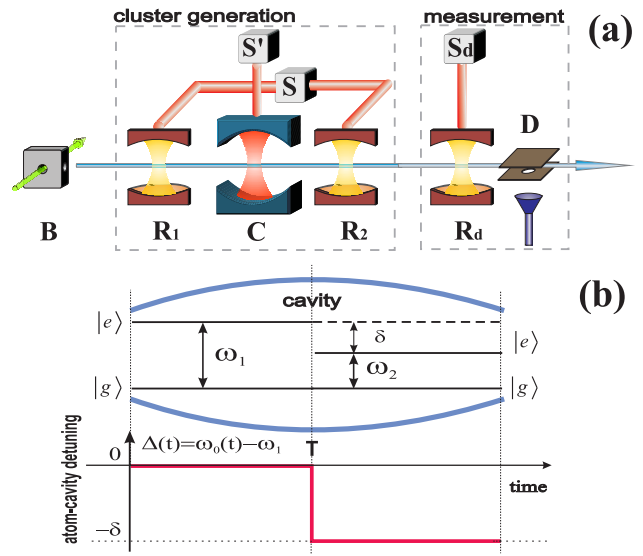}
\caption{(Color online) (a) Schematic set-up of a microwave cavity
experiment in which a chain of Rydberg atoms is emitted from a
source $B$ and then passes through a Ramsey zone $R_1$, a cavity
$C$, the Ramsey zones $R_2$ and $R_d$, and until the atoms are
ionized and observed finally at the detector $D$. The classical
fields in the Ramsey zones are generated by the microwave sources
$S$, $S^\prime$ and $S_d$. (b) Temporal matching of the $e
\leftrightarrow g$ atomic transition frequency ($\omega_0$) to
either the frequency $\omega_1$ of the first cavity mode or the
frequency $\omega_2$ of the second mode in the course of the
resonant atom-cavity interaction. Apart from the matching of the
atomic frequency (upper half), the lower part of this figure
displays the time dependence of the atom-cavity detuning $\Delta(t)
= \omega_{0}(t) - \omega_{1}$, implying a step-wise change from the
resonant $A-M_1$ interaction regime $(t < T)$ to the resonant
$A-M_2$ regime $(t > T)$. See text for further discussions.}
\label{fig:1}
\end{center}
\end{figure}

\section{Generation of cluster states by using bimodal cavities}

The resonant atom-cavity interaction is perhaps the simplest regime
that can be used to entangle the circular excited states of a
Rydberg atom, in a well-controlled way, with the quantized field
states of a cavity. For a sufficiently high quality (factor) of the
cavity mirrors, this regime implies a strong atom-field coupling for
which the dissipation of energy from the cavity becomes negligible
during the interaction period. Indeed, a negligible dissipation is
crucial for the engineering of multipartite entangled states between
atomic qubits, if the cavity mediates this entanglement formation.
Apart from the quality of the cavity, the correct matching of the
atomic transition frequency to the resonant frequency of the cavity
mode(s), the so-called detuning, is also important in order to
realize a resonant interaction between the atom and the cavity.

In the following, let us adopt here the language of the Haroche
group \cite{haroche, haroche1} for describing the cavity QED
experiments and to specify the states of the atoms and the cavity.
In the experiments of Haroche and coworkers, rubidium atoms are
prepared to occupy one of the three Rydberg levels with principal
quantum numbers 51, 50, or 49, and which are referred to as excited
state $\ket{e}$, ground state $\ket{g}$, and auxiliary state
$\ket{a}$, respectively. Owing to the particular design of the
microwave cavity, however, only the states $\ket{e}$ and $\ket{g}$
can be involved in the atom-cavity interaction because only the $e
\leftrightarrow g$ transition frequency of an rubidium atom can be
tuned to the frequency of the cavity mode(s). The classical
microwave field from the sources $S$, $S^\prime$ and $S_d$
[cf.~Fig.~\ref{fig:1}(a)], in contrast, can be adapted to drive the
$e \leftrightarrow g$ or $g \leftrightarrow a$ transitions and are
utilized to generate or manipulate the superposition between these
atomic states when the atom interacts with the microwave field.

The (time) evolution of an atom coupled to single-mode cavity is
described by the Jaynes-Cummings Hamiltonian \cite{jc} ($\hbar = 1$)
\begin{equation}\label{ham}
H = \omega_{0} S_{z} - \im \frac{\Omega}{2}
        \left( S_{+} a_1 - a_1^{+} S_{-} \right)
      + \omega_1 \left(a_1^{+} a_1 + \frac{1}{2} \right),
\end{equation}
where $\omega_{0}$ is the atomic $e \leftrightarrow g$ transition
frequency, $\omega_1$ the frequency of the cavity field, and
$\Omega$ the atom-field coupling frequency. In this Hamiltonian,
moreover, $a_1$  and $a_1^+$ denote the annihilation and creation
operators for a photon in the cavity, that act upon the Fock states
$\ket{n}$, while $S_{-}$ and $S_{+}$ are the atomic spin lowering
and raising operators that act upon the excitation states $\ket{e}$
and $\ket{g}$, and which are the eigenstates of $S_z \equiv \sigma_z
/2$ with eigenvalues $+1/2$ and $-1/2$, respectively. Due to the
Hamiltonian (\ref{ham}), the overall atom-field state evolves during
the resonant interaction, e.g., for a zero detuning ($0 = \omega_0 -
\omega_1$), as
\begin{subequations}\label{e1}
\begin{eqnarray}
\label{e1a} \ket{e,0} & \rightarrow &
          \cos{\left(\Omega t /2 \right)} \,\ket{e,0} +
          \sin{\left(\Omega t /2 \right)} \,\ket{g,1}, \\[0.1cm]
\label{e1b} \ket{g,1} & \rightarrow &
          \cos{\left(\Omega t /2 \right)} \,\ket{g,1} -
          \sin{\left(\Omega t /2 \right)} \,\ket{e,0} \, ,
\end{eqnarray}
\end{subequations}
i.e.\ with a time evolution that is known also as Rabi rotation. In
this (Rabi) picture, $t$ is the effective atom-cavity interaction
time in the laboratory and $\Omega \, t$ is the corresponding angle
of rotation. Note that neither the state $\ket{e,1}$ nor $\ket{g,0}$
appears in the time evolution (\ref{e1}) in line with our physical
intuition that the photon energy is stored either by the atom or the
cavity but should not occur twice in the system.

In contrast to a single-mode cavity, bimodal cavities support two
non-degenerate modes of light with (usually) orthogonal
polarization. Since the frequencies of these cavity modes are fixed
by the design and geometry of the cavity mirrors, the atomic
frequency $e \leftrightarrow g$ need to be (de-)tuned in the course
of the interaction such that the atom can interact resonantly with
either the first or the second cavity mode. In the language of
quantum information, the additional cavity mode gives rise to
another photonic qubit that may interact independently with the
atomic qubits, while they are passing through the cavity.

From the first experiments with bimodal cavities \cite{pra64, prl75,
prl92}, their use has been found an important step towards the
generation and control of entangled states. Indeed, a number of
proposals \cite{pra68, jmo51, pra67, pra77, pla339, pra78} has been
made in the literature in order to exploit further capabilities of
bimodal cavities concerning, for example, the coherent manipulation
of complex quantum states or for performing fundamental tests on
quantum theory. Below, we shall denote the two cavity modes by
$M_{1}$ and $M_{2}$ and assume that they are associated with the
frequencies $\omega_1$ and $\omega_2$, such that $\omega_1 -
\omega_2 \equiv \delta > 0$. For the cavity utilized in the
experiments by Rauschenbeutel and coworkers \cite{pra64}, in
particular, a frequency splitting of $\delta / 2\pi = 128.3$ KHz was
realized. Owing to this fixed splitting in the frequency of the
field modes, we refer to the detuning of the atomic transition
frequency with regard to the first cavity mode frequency: $\Delta
(t) \: \equiv \: \omega_0(t) \,-\, \omega_1$, briefly as the
atom-cavity detuning.

An entanglement of a Rydberg atom with the photon field of the
cavity is then achieved in a controlled way by tuning the $e
\leftrightarrow g$ transition frequency as function of time, so that
it is in resonance with either one or the other cavity mode, while
the atom passes through the cavity. For a sufficiently fast switch
of the detuning $\Delta(t)$, i.e.\ of the atomic frequency between
the two modes of the cavity, a resonant interaction (regime) is
realized with either mode $M_{1}$ for $\Delta(t < T) = 0$
\textit{or} with the mode $M_{2}$ for $\Delta(t > T) = -\delta$,
cf.\ Figure~\ref{fig:1}(b), and where usually a step-wise change
from the $A - M_1$ to the $A - M_2$ interaction is assumed. In the
experiments by Haroche and coworkers, the detuning is changed by
applying a well adjusted time-varying electric field across the gap
between the cavity mirrors, so that the required (Stark) shift of
the atomic $e \leftrightarrow g$ transition frequency is obtained.
Instead of the instantaneous (step-like) change of the atom-cavity
detuning, however, only a--more or less--smooth switch can be
realized for the detuning of the atomic frequency within the finite
time of $\simeq 1 \, \mu s$. In practice, such a finite switch is
not completely negligible and may affect the evolution of the cavity
states \cite{jpb}. In the present work, however, we shall not
consider the effects of this finite switching time but assume a
step-wise change in the detuning as indicated in the lower part of
Fig.~\ref{fig:1}(b).

Let us mention here, moreover, that the atom can interact resonantly
at any given time only with one of the modes, while the second mode
is then frozen out from the interaction because of the (large)
splitting $\delta$ between the two cavity modes. Therefore, the
overall $A - M_{1} - M_{2}$ time evolution of the atom-cavity state
can be separated into two independent parts: the evolution that
occurs due to the $A - M_{1}$ resonant interaction as displayed in
Eq.~(\ref{e1}), and the evolution due to $A - M_{2}$
\begin{subequations}\label{e2}
\begin{eqnarray}
\ket{e,\bar{0}} & \rightarrow &
                 \cos{\left(\Omega t /2 \right)} \ket{e,\bar{0}} +
                 \im \sin{\left(\Omega t /2 \right)} \ket{g,\bar{1}},
                 \quad \label{e2a}
\\[0.1cm]
\ket{g,\bar{1}} & \rightarrow &
                \cos{\left(\Omega t /2 \right)} \ket{g,\bar{1}} +
                \im \sin{\left(\Omega t /2 \right)} \ket{e,\bar{0}} \, .
                \quad \label{e2b}
\end{eqnarray}
\end{subequations}
In the evolution above, the states $\ket{\bar{0}}$ and
$\ket{\bar{1}}$ hereby refer to the Fock states of the cavity mode
$M_2$ and the $\im$ factor arises due to orthogonal polarization of
the mode $M_2$ with respect to the mode $M_1$.

With this short reminder on the Jaynes-Cummings Hamiltonian and the
(atom-cavity) interactions in a bimodal cavity, we are now prepared
to present all the steps that are necessary in order to generate one
and two-dimensional cluster states for a chain of Rydberg atoms
which crosses the cavity set-up.

\begin{figure}
\begin{center}
\includegraphics[width=0.46\textwidth]{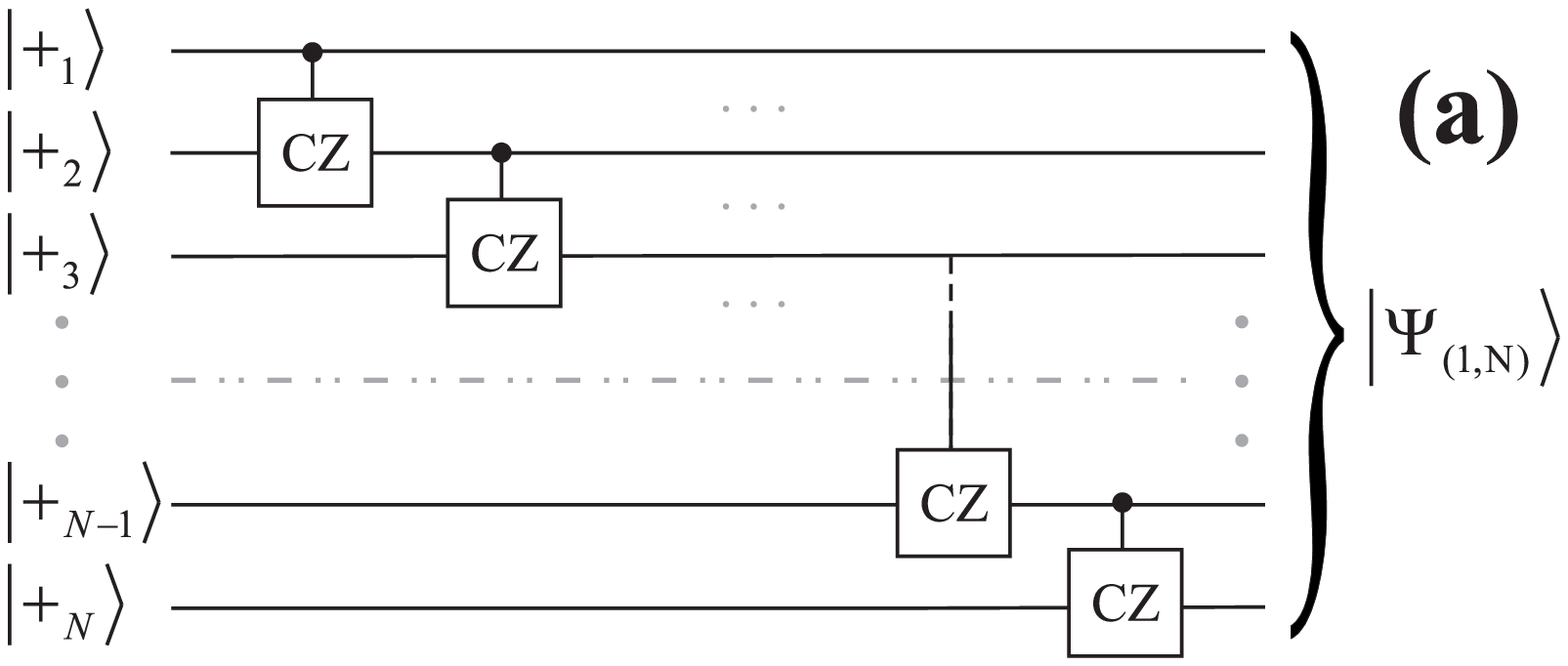} \\
\vspace{0.5cm}
\includegraphics[width=0.46\textwidth]{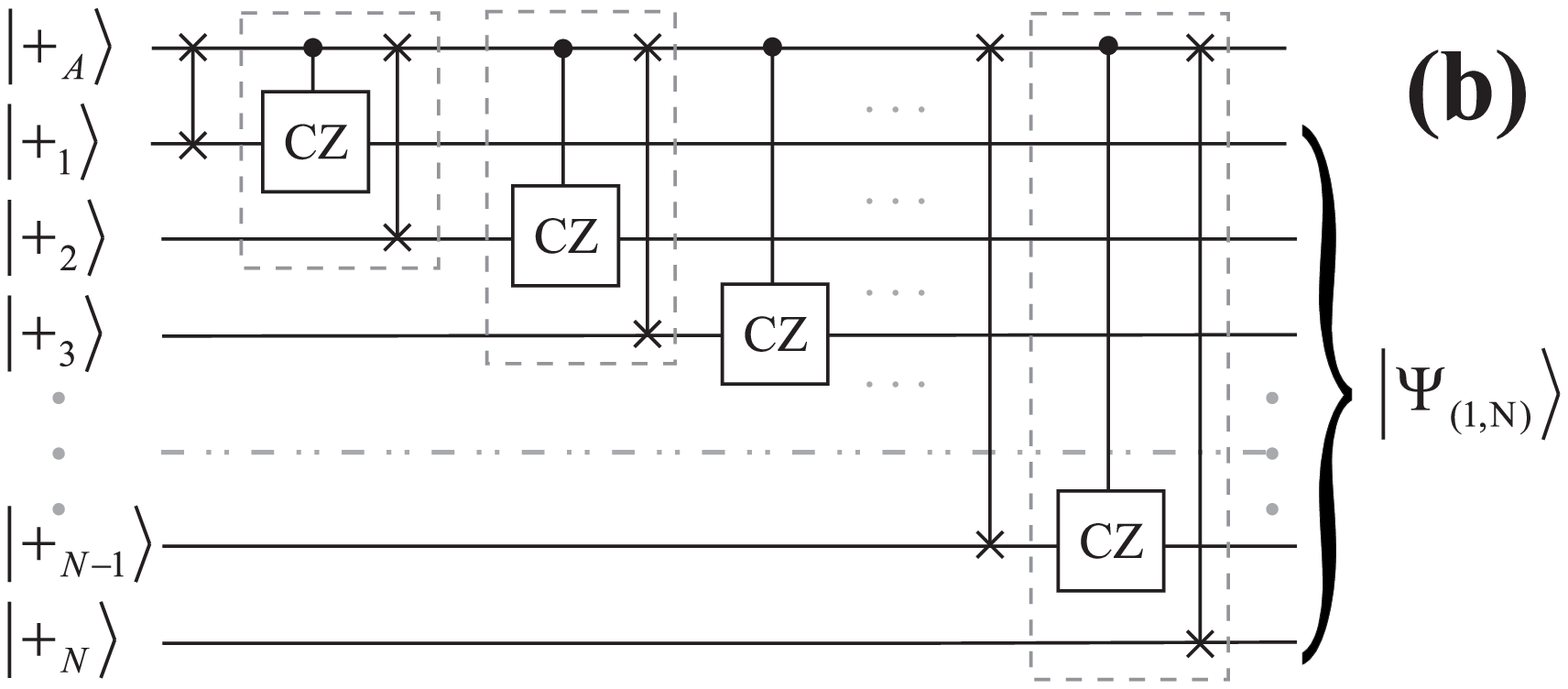}
\caption{(a) Quantum circuit for the generation of a linear cluster
state between $N$ uncorrelated qubits. Each qubit is initially
prepared in the $\ket{+}$ state, and one controlled-z gate is
applied then subsequently to any two neighboring qubits. (b)
Alternative quantum circuit for the linear cluster state generation,
in which the controlled-z operation is successively applied to the
ancilla qubit $A$ and qubit $k$, and followed by the swapping of the
ancilla state with the qubit $k+1$.}
\label{fig:2}
\end{center}
\end{figure}

\subsection{Linear Cluster State}

In this subsection, we first explain the generation of a linear $1
\times N$ cluster state for a chain of atoms, and for which only a
single cavity mode is required. This scheme for the generation of
linear states was first suggested by Sch\"{o}n and coworkers
\cite{mps} and will be adapted here for the cavity set-up as
displayed in  Fig.~\ref{fig:1}(a).

The linear cluster state is defined  as \cite{BR1}
\begin{equation}\label{lin-clust-def}
\ket{\Psi_{(1,N)}} = \frac{1}{2^{N/2}}
\overset{N}{\underset{i=1}{\otimes}} \left( \ket{0_i} + \ket{1_i} \,
\Theta_{i+1} \right),
\end{equation}
where $\Theta_k \equiv \ket{0_k}\bra{0_k} - \ket{1_k}\bra{1_k}$ acts
on the $k$-th qubit and $\Theta_{N+1} \equiv 1$. Alternatively one
can define the linear cluster state also as a lattice of $N$ qubits,
where the nodes refer to the qubits that are initialized
(altogether) in the product state $\ket{+_1} \otimes \ldots \otimes
\ket{+_N}$, and where the edges of the lattice refer to the
two-qubit controlled-z gates (\ref{cz-def}) that are applied between
neighboring nodes. According to this latter definition,
Fig.~\ref{fig:2}(a) displays the successive interactions which are
necessary to construct the linear cluster state for $N$ initially
uncorrelated qubits. Instead of applying the controlled-z gate to
each pair of neighboring qubits $k$ and $k+1$, however, we can apply
this two-qubit gate to the ancilla qubit $A$ and the ordinary qubit
$k$, and then swap the state of $A$ with the qubit $k+1$ as
displayed in Fig.~\ref{fig:2}(b). Note that, in this circuit, we
have inserted one additional swap gate between the ancilla and the
first atom which has no effect on the output cluster state since the
ancilla qubit is prepared initially also in the state $\ket{+}$.

Below we shall associate the ancilla qubit with the cavity mode
$M_1$ and the ordinary qubits with the (two-level) Rydberg atoms.
According to the second scheme from Fig.~\ref{fig:2}(b), this
identification implies that the atoms pass sequentially through the
cavity and that only one atom couples to the cavity at a time, which
fits nicely to our cavity set-up displayed in Fig.~\ref{fig:1}(a).
As seen from Fig.~\ref{fig:2}(b), moreover, only two types of
unitary gates have to be performed between the cavity mode and the
atoms that cross through the cavity, namely, (i) the swap gate
followed by the controlled-z gate for atoms $A_1 \ldots A_{N-1}$ and
(ii) the swap gate for the atom $A_N$. Before the atom-cavity
interaction starts, moreover, each atom must be prepared in the
superposition $\ket{+} \equiv (\ket{e} + \ket{g})/\sqrt{2}$. In the
set-up above, this initial superposition is achieved by first
exciting the atoms from the source $B$ into the state $\ket{e}$ and
then by applying the rotation
\begin{equation}\label{rot1}
\ket{e} \rightarrow \frac{1}{\sqrt{2}} \left( \ket{e} + \ket{g}
\right)
\end{equation}
just before the atom enters the cavity. As explained below, the
rotation (\ref{rot1}) can be realized efficiently by means of the
Ramsey zone $R_1$.

\begin{figure}
\begin{center}
\includegraphics[width=0.46\textwidth]{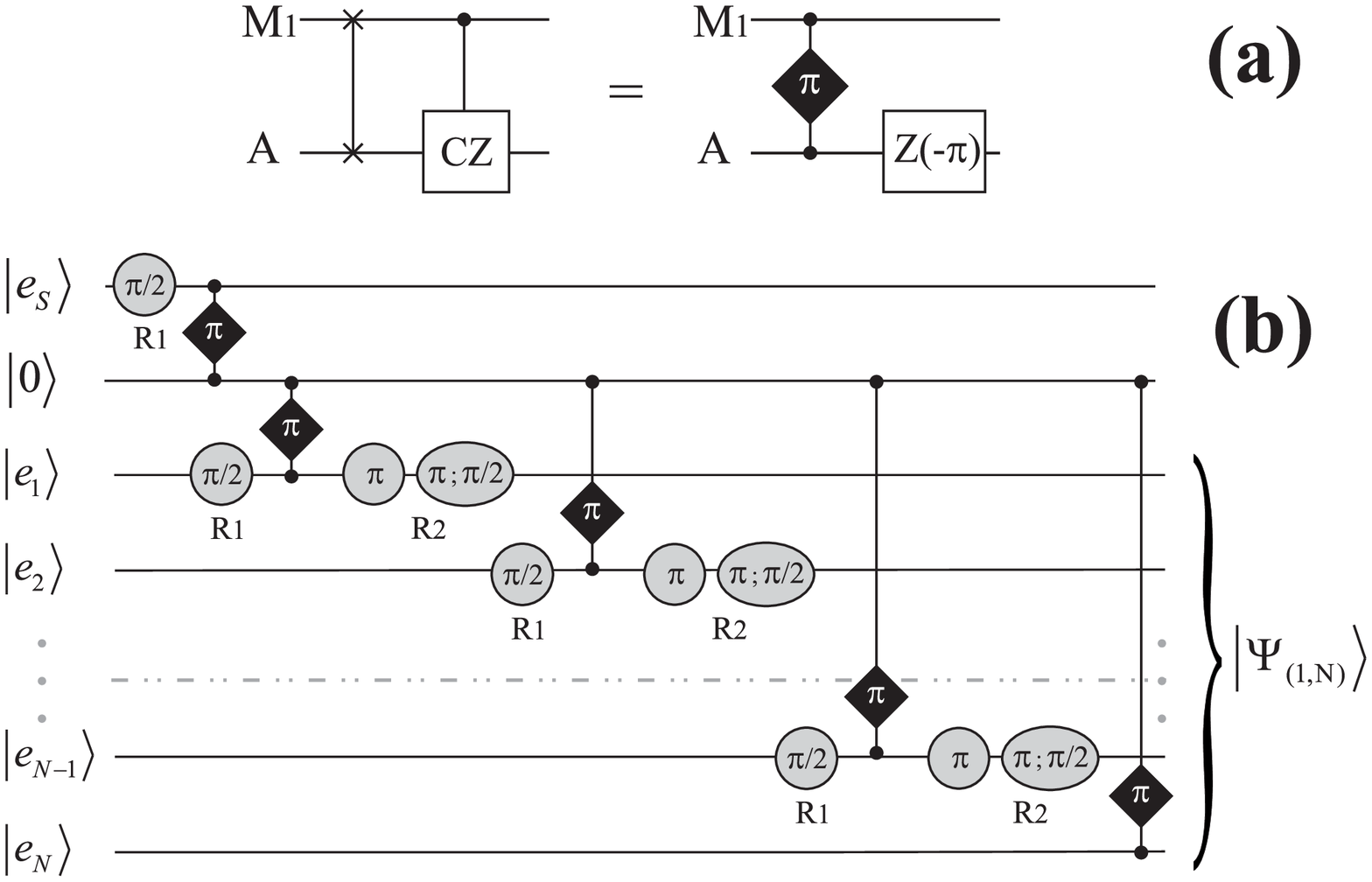} \\
\vspace{0.5cm}
\includegraphics[width=0.46\textwidth]{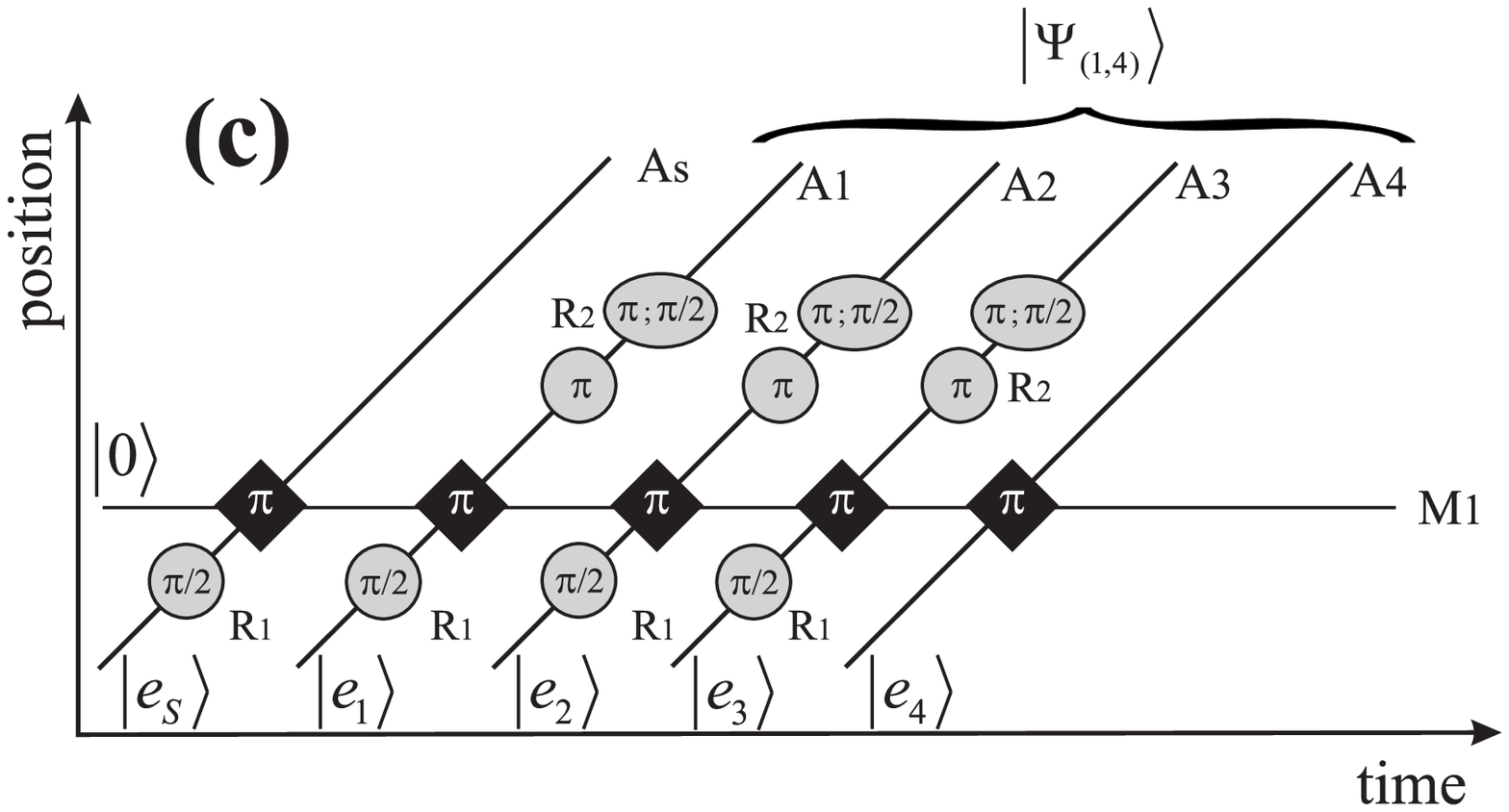}
\caption{(a) Two equivalent circuits that follow from relation
(\ref{rel1}) after multiplying it by $\left[ Z(\pi) \otimes I
\right]^{-1}$ and where the factor $(-\im)$ has been omitted for
brevity. (b) Quantum circuit for the generation of a linear cluster
state that is encoded into a chain of $N$ Rydberg atoms passing
through the cavity. (c) Temporal sequence that corresponds to the
above circuit for the case $N=4$. The pictograms and notation in
these figures are explained in the text.}
\label{fig:3}
\end{center}
\end{figure}

The evolution of the atom-cavity state due to the resonant
interaction of the atom with the cavity mode $M_1$ $(\Delta = 0)$ is
given by Eqs.~(\ref{e1}) which, for a Rabi rotation $\Omega t =
\pi$, is equivalent to the modified swap gate
\begin{equation}\label{m-swap}
U^{\rm m-swap} = \left(
\begin{array}{cccc}
1          & 0     & 0     & 0  \\
0          & 0     & -1    & 0  \\
0          & 1     & 0     & 0  \\
0          & 0     & 0     & 1
\end{array}
\right) \, ,
\end{equation}
expressed in the basis $\{ \ket{g,0}, \ket{g,1}, \ket{e,0},
\ket{e,1} \}$. In contrast to the conventional swap gate (which has
no minus sign), we shall therefore refer to this two-qubit operation
as m-swap gate below. Following the work by Sch\"{o}n and coworkers,
we can express the m-swap gate (\ref{m-swap}) also in the form
\begin{equation}\label{rel1}
U^{\rm m-swap} = (- \im) \; U^{\rm swap} \cdot U^{\rm cz} \cdot
\left[ Z(\pi) \otimes I \right],
\end{equation}
where
\begin{equation}\label{gates}
U^{\rm swap} = \left(
\begin{array}{cccc}
1          & 0     & 0     & 0  \\
0          & 0     & 1     & 0  \\
0          & 1     & 0     & 0  \\
0          & 0     & 0     & 1
\end{array}
\right); \;
U^{\rm cz} = \left(
\begin{array}{cccc}
1          & 0     & 0     & 0  \\
0          & 1     & 0     & 0  \\
0          & 0     & 1     & 0  \\
0          & 0     & 0     & -1
\end{array}
\right)
\end{equation}
are the swap and controlled-z gates taken in the same basis as the
matrix (\ref{m-swap}), and where $Z(\theta) \equiv e^{-i \sigma_z
\theta /2}$ denotes the standard rotation operator with regard to
the (quantization) $z-$axis that acts upon the atomic state.
Eq.~(\ref{rel1}) implies that the m-swap gate is equivalent (up to a
constant phase) to the swap gate followed by the controlled-z gate
together with a local rotation of the atomic state. In order to
realize only the swap gate followed by the controlled-z gate as
required by our scheme [see Fig.~\ref{fig:2}(b)], therefore, the
m-swap gate (atom-cavity $\pi$ rotation) should be followed by the
local rotation $Z^{-1}(\pi) = Z(-\pi)$ on the atom as depicted
graphically in Fig.~\ref{fig:3}(a).

Up to this point, we just summarized a scheme that enables one to
generate a linear $1 \times N$ cluster state by sending a chain of
$N$ uncorrelated atoms through the cavity in such a way that only
one atom couples to the cavity mode at a time. Specifically, we have
shown that each atom is incorporated into the cluster state by
performing the superposition (\ref{rot1}) followed by a Rabi
rotation $\Omega t \:=\: \pi$ of the atom-cavity system and
finalized by a $Z(-\pi)$ rotation of the atomic state. In order to
fully adapt this scheme for our cavity set-up given in
Fig.~\ref{fig:1}(a), it is necessary to express these atomic
rotations in terms of those classical (field) pulses that can be
generated by means of the microwave source $S$. These pulses have to
be applied when the atom passes through the Ramsey zones, either in
front ($R_1$) and/or behind ($R_2$) of the cavity.

The interaction of a Rydberg atom with a (classical) microwave field
gives rise to the unitary transformation of the atomic state
\cite{haroche, haroche1}
\begin{equation}\label{ramsey}
R(\phi, \varphi) = \left(
\begin{array}{cc}
\cos \left( \phi /2 \right)  &  -\sin \left( \phi /2 \right) e^{-\im \varphi} \\
\sin \left( \phi /2 \right) e^{\im \varphi}  &  \cos \left( \phi /2
\right)
\end{array}
\right)
\end{equation}
expressed in the basis $\{ \ket{g}, \ket{e} \}$. In this rotation
matrix, the angle $\phi$ is proportional to the duration of the
microwave pulse, and an additional phase $e^{\im \varphi}$ is
accumulated whenever the microwave field frequency is slightly
detuned from the atomic $e \leftrightarrow g$ transition frequency.
In the literature, such an interaction is often called a Ramsey
pulse. Using the unitary matrix (\ref{ramsey}), it can be easily
seen that one resonant $\pi/2$ Ramsey pulse gives rise to the
rotation (\ref{rot1}), and that the rotation around the $z-$axis
\begin{equation}\label{rot2}
Z(-\pi) = R(\pi,0) \cdot R \left( \pi, \pi/2 \right),
\end{equation}
is obtained by applying two $\pi$ Ramsey pulses successively, where
one is resonant and another one detuned by $\pi/2$. In typical
cavity QED experiments, a single $\pi$ Ramsey pulse takes about
$2\,\mu s$ and thus implies that the atom is still inside of the
Ramsey plates when the required rotation of the atomic state has
been completed. Therefore, after a short time delay, it is possible
to apply one additional (detuned) Ramsey pulse upon the same atom
and within the same Ramsey zone. This enables one to realize the
rotation (\ref{rot1}) while the atom crosses the Ramsey zone $R_1$
in front of the cavity, and the $Z(-\pi)$ rotation by the Ramsey
zone $R_2$ just behind the cavity [see Fig.~\ref{fig:1}(a)].

With this analysis, we now have all ingredients available to
generate linear cluster states within our cavity set-up given in
Fig.~\ref{fig:1}(a). The equivalent quantum circuit for this scheme
is displayed in Fig.~\ref{fig:3}(b), in which the atom-cavity
interactions are depicted by black diamonds (including the Rabi
rotation angle), while the Ramsey pulses $R(\phi, \varphi)$ are
shown as gray circles. For these Ramsey pulses, we also display the
interaction time in units of rotation angle $\phi$ and the phase
$\varphi$ if it is non-zero. In addition, the letters $R_1$ or $R_2$
are utilized in order to denote the Ramsey zones in front or behind
the cavity. Note that, in order to prepare the cavity mode in the
$\ket{+}$ state, we made use of an auxiliary atom $A_s$ that is
initialized in the excited state and which crosses the cavity before
the chain of atoms arrives. This auxiliary atom interacts for a
$\pi/2$ Ramsey pulse with the microwave field $R_1$ and then for a
$\pi$ Rabi pulse with the cavity. According to Eq.~(\ref{rot1}) and
Eqs.~(\ref{e1}), the initially empty cavity field is then set to the
$\ket{+}$ state, while the auxiliary atom is factorized out in its
ground state. Let us also note that the last swap gate between the
cavity and the $N$-th atom is replaced by the m-swap gate
(\ref{m-swap}), which simply maps the cavity state $\ket{0}$ upon
the atomic ground state $\ket{g}$ and the cavity state $\ket{1}$
upon the excited state $\ket{e}$. This replacement finally
factorizes out the cavity state from the atomic cluster state.

Fig.~\ref{fig:3}(c) shows the particular temporal sequence that can
be applied for the generation of the four-qubit cluster state
\begin{eqnarray}\label{lin-clust}
\ket{\Psi_{(1,4)}} = & \frac{1}{2} & \left( \ket{g_1, +_2, g_3, +_4}
                         + \ket{g_1, -_2, e_3, -_4} \right. \\
                     &&  \left. + \ket{e_1, -_2, g_3, +_4} +
                         \ket{e_1, +_2, e_3, -_4}  \right)  \notag
\end{eqnarray}
for a chain of four atoms, where the notation $\ket{\pm} = (\ket{g}
\pm \ket{e}) / \sqrt{2}$ has been used and the factorized state
$\ket{g_s,1}$ of the auxiliary atom and cavity mode is not displayed
for brevity. This (temporal) sequence is just another way of
presenting the set of Rabi and Ramsey pulses which was introduced
originally by Haroche and coworkers in order to depict graphically
the unitary evolution of the atom-field state in the framework of
cavity QED. It is straightforward to check that one obtains (up to a
constant phase) the state (\ref{lin-clust}) if all the unitary
transformations from this sequence are properly evaluated. It is
also obvious that the state (\ref{lin-clust}) is equivalent to the
state (\ref{lin-clust-def}) for $N=4$ by considering the assignments
\begin{eqnarray}
\begin{array}{llll}
   \ket{g_1} = \ket{0_1}, & \ket{g_2} = \ket{0_2}, &
   \ket{g_3} = \ket{0_3}, & \ket{g_4} = \ket{0_4}, \\[0.1cm]
   \ket{e_1} = \ket{1_1}, & \ket{e_2} = \ket{1_2}, &
   \ket{e_3} = \ket{1_3}, & \ket{e_4} = \ket{1_4}.
   \end{array}  \notag
\end{eqnarray}
In this subsection, we have shown that each atom from a chain of $N$
uncorrelated atoms is incorporated into the linear cluster state by
performing a Rabi $\pi$ rotation and (if required) three Ramsey
pulses: one applied before and two behind the cavity.

\subsection{$2 \times N$ Cluster State}
\begin{figure}
\begin{center}
\includegraphics[width=0.46\textwidth]{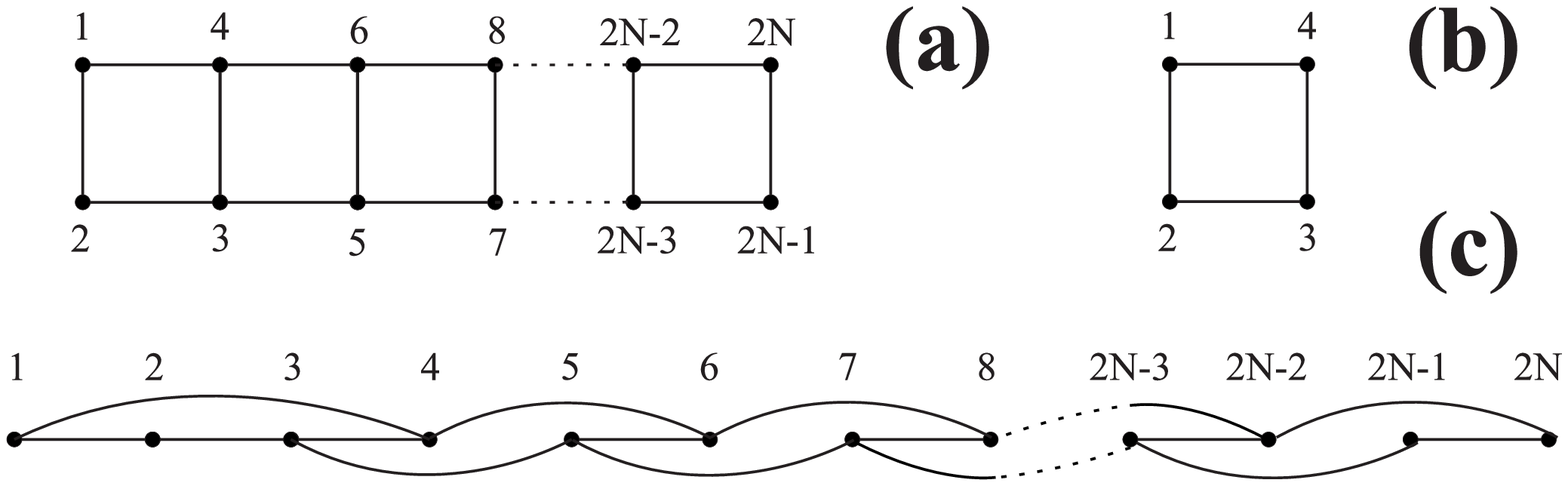} \\
\vspace{0.5cm}
\includegraphics[width=0.46\textwidth]{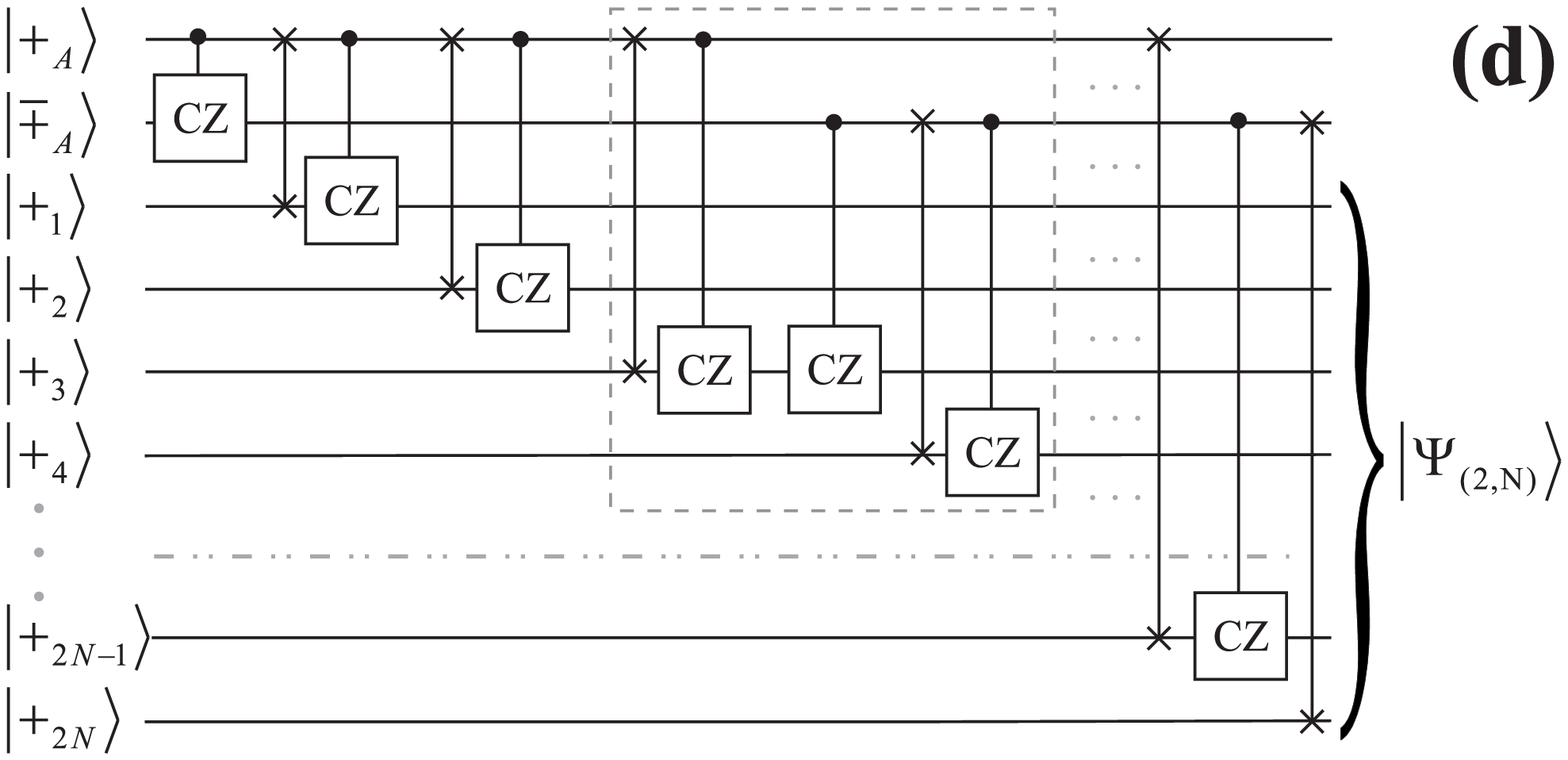}
\caption{(a) Two-dimensional $2 \times N$ cluster state. (b) Box
state (\ref{box-def}) that is the simplest two-dimensional cluster
state. (c) Definition of the edges for a chain of $2N$ atoms
(nodes), such that an effective two-dimensional $2 \times N$ cluster
state is produced. The labels of the nodes in Figure (a) correspond
to the serial numbers of the atoms inside the chain. (d) Quantum
circuit for the generation of the $\ket{\Psi_{(2,N)}}$ cluster state
between $2N$ initially uncorrelated qubits, and for which two
ancilla qubits are utilized. In this circuit, the controlled-z gates
(edges) are applied according to Figure (c).}
\label{fig:4}
\end{center}
\end{figure}
\begin{figure*}
\begin{center}
\includegraphics[width=0.75\textwidth]{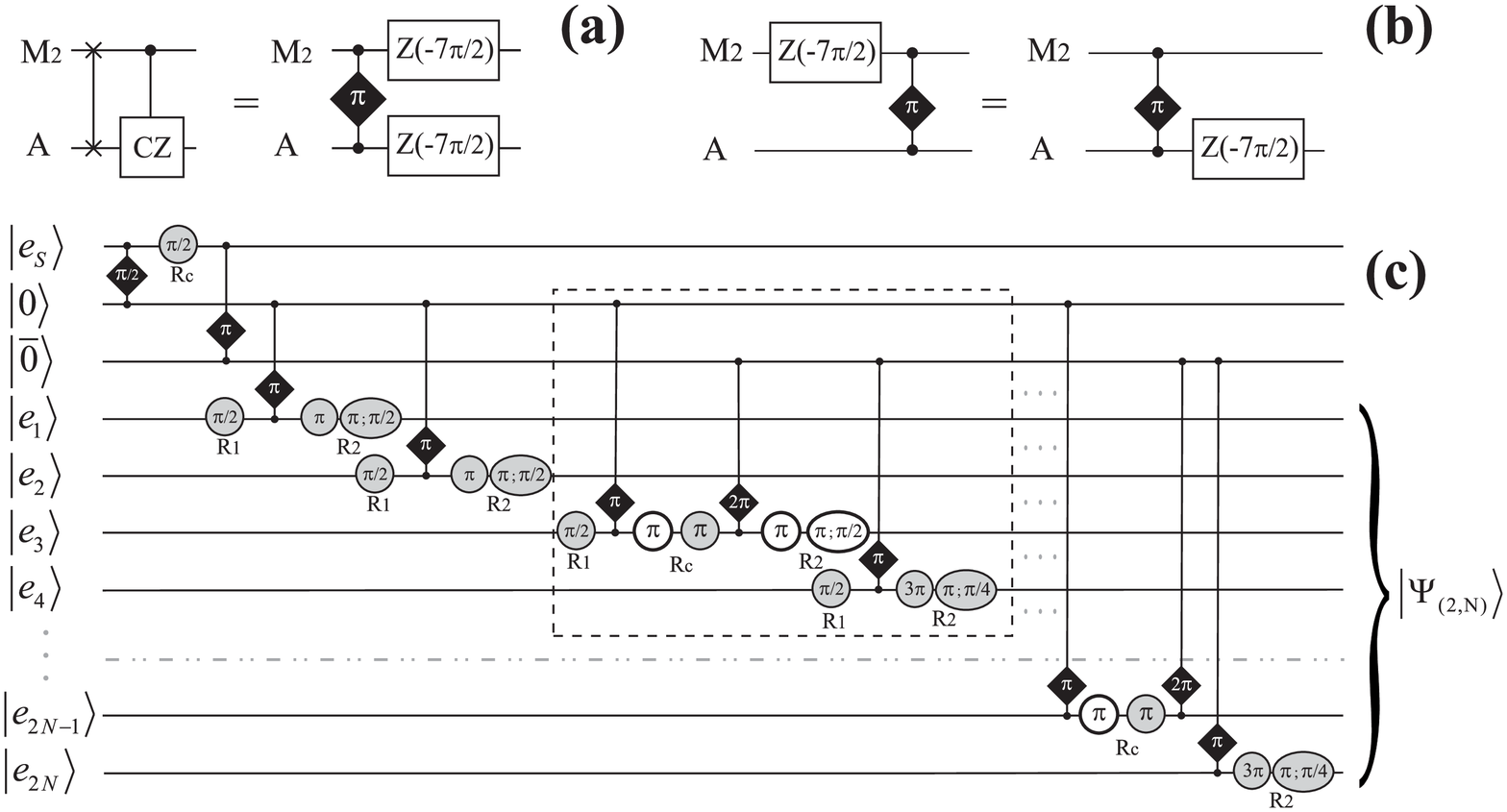}
\caption{(a) Two equivalent circuits that follow from relation
(\ref{rel2}) after multiplying it by $\left[ Z(7\pi/2) \otimes
Z(7\pi/2) \right]^{-1}$ and where the factor $\im$ has been omitted
for brevity. (b) Recipe that follow from relation (\ref{rel3}) and
allows for replacing of the cavity state rotation during the step
$k$ by the same rotation of the atomic state for the step $k+1$. (c)
Quantum circuit for the generation of the $2 \times N$ cluster state
that is associated with a chain of $2N$ Rydberg atoms passing
through the cavity. The new white-circled pictogram and $R_c$
notation are explained in the text. Note that the gates inside the
dashed boxed as well as the rotations of the last qubit $A_{2N}$
inside the Ramsey zone $R_2$ should be omitted for $N = 2$.}
\label{fig:5}
\end{center}
\end{figure*}

A linear cluster state alone is not sufficient for universal one-way
quantum computations since its structure does not enable one to
encode quantum gates that act upon several qubits. In this work, we
therefore introduce a scheme that generates two-dimensional cluster
states $\ket{\Psi_{(2,N)}}$, i.e.\ states that form a
two-dimensional lattice of $2 \times N$ qubits. In this lattice, the
qubits are initialized in the state $\ket{+_1, \ldots +_{2N}}$, and
a controlled-z gate is applied for all edges that connect
neighboring nodes, cf.~Fig.~\ref{fig:4}(a). In fact, the concept of
cluster states is neither restricted to a rectangular pattern of
nodes not that only nearest neighbours could be connected with each
other by a controlled-z operation. In the present work, however, we
shall confine ourselves to two-dimensional clusters with rectangular
geometry as displayed in Fig.~\ref{fig:4}(a).

The simplest example of such a two-dimensional cluster state is the
(so-called) box state \cite{nat434}
\begin{eqnarray}\label{box-def}
\ket{\Psi_{(2,2)}} = & \frac{1}{2} & \left( \ket{0_1, +_2, 0_3, +_4}
                         + \ket{0_1, -_2, 1_3, -_4} \right. \\
                     &&  \left. + \ket{1_1, -_2, 0_3, -_4} +
                         \ket{1_1, +_2, 1_3, +_4}  \right), \notag
\end{eqnarray}
with $\ket{\pm} = (\ket{0} \pm \ket{1}) / \sqrt{2}$. Note that this
four-qubit cluster state could be obtained alternatively from the
linear cluster state (\ref{lin-clust}) by means of the
transformations
\begin{equation}\label{transf3}
\ket{g_1, \pm_4} \rightarrow \ket{g_1, \pm_4}, \quad \ket{e_1,
\pm_4} \rightarrow \ket{e_1, \mp_4},
\end{equation}
and together with a change in the notation of the (atomic) qubits:
$\ket{g} \rightarrow \ket{0}$ and $\ket{e} \rightarrow \ket{1}$.
These two transformations can be realized by applying the $A_1 -
A_4$ controlled-z gate that gives rise to the box configuration of
Fig.~\ref{fig:4}(b) in which the first and the last qubits of the
linear chain are now connected by an edge.

Following our set-up in Fig.~\ref{fig:1}(a), however, we made the
(realistic) assumption from the very beginning that only a single
chain of atoms is produced by the atomic source and sent into the
cavity. For this reason, we need to consider a different procedure
(if compared with the linear cluster states) for defining the edges
between the nodes associated to a chain of $2N$ atoms such that on
the output, after all the atoms have crossed the cavity, an
effective two-dimensional cluster state is generated. As mentioned
above, bimodal cavities are designed in a way so they support two
independent modes of the photon field. This makes it possible to
implement schemes in which two (photonic) ancilla qubits are
utilized due to the two cavity modes $M_1$ and $M_2$. This enables
us to generate the $2N$-partite entangled state displayed in
Fig.~\ref{fig:4}(c) and that represents the two-dimensional $2
\times N$ cluster state upon the assignment of the atomic positions
in a chain of $2N$ atoms to the two-dimensional cluster state as
shown in Fig.~\ref{fig:4}(a). The quantum circuit that accomplishes
this task is displayed in Fig.~\ref{fig:4}(d), in which the gates
that are placed inside of the dashed boxed area need to be repeated
$N-3$ times. Apart from the $A-M_1$ unitary gate, which we
introduced in the previous subsection, this circuit contains three
additional gates. Two of them act upon the $A-M_2$ system: (i) the
swap gate followed by the controlled-z gate, and (ii) a single
controlled-z gate. The third gate is the controlled-z gate that acts
upon the $M_1-M_2$ system. We shall discuss all of these operations
now in turn.

Eqs.~(\ref{e2}) display the evolution of the atom-cavity state for a
resonant interaction of an atom with the cavity mode $M_2$ $(\Delta
= -\delta)$ which, for a Rabi rotation $\Omega t = \pi$, gives rise
to the i-swap gate \cite{pra67a}
\begin{equation}\label{i-swap}
U^{\rm i-swap} = \left(
\begin{array}{cccc}
1          & 0     & 0     & 0  \\
0          & 0     & \im   & 0  \\
0          & \im   & 0     & 0  \\
0          & 0     & 0     & 1
\end{array}
\right),
\end{equation}
expressed in the basis $\{ \ket{g,\bar{0}}, \ket{g,\bar{1}},
\ket{e,\bar{0}}, \ket{e,\bar{1}} \}$. Similar as in
Eq.~(\ref{rel1}), we re-write this i-swap gate as
\begin{equation}\label{rel2}
U^{i-swap} = \im \; U^{swap} \cdot U^{cz} \cdot \left[ Z
\left(\frac{7\pi}{2} \right) \otimes Z \left(\frac{7\pi}{2} \right)
\right] \, .
\end{equation}
Therefore, the swap gate followed by the controlled-z gate is
performed by applying the $\pi$ Rabi pulse followed by the local
$Z^{-1}(7\pi/2) = Z(-7\pi/2)$ rotations of both, the atom and the
cavity states, see also Fig.~\ref{fig:5}(a). A rotation of the
atomic state can be easily implemented by utilizing two Ramsey
pulses
\begin{equation}\label{rot3}
Z(-7\pi/2) = R(3\pi,0) \cdot R \left( \pi, \pi/4 \right)  ,
\end{equation}
separated from each other by a short time delay [cf.\
Eq.~(\ref{rot2})]. Unfortunately, however, it is not easy to perform
local operations on the cavity state. Therefore, we shall make use
of the relation
\begin{equation}\label{rel3}
\left[ I \otimes Z(\theta) \right] \cdot U^{i-swap} = U^{i-swap}
\cdot \left[ Z(\theta) \otimes I \right] \, ,
\end{equation}
which provides us with a hint, namely, we can replace the rotation
of the cavity state during the step $k$ by the same rotation of the
atomic state in the step $k+1$. The diagram for this replacement is
displayed in Fig.~\ref{fig:5}(b) and will be utilized in our further
discussion.

Apart from the two basis states $\ket{e}$ and $\ket{g}$ of the
atomic qubit, as applied in the atom-cavity interaction above, the
Rydberg electron can populate also the auxiliary state $\ket{a}$
below of the (ground) state $\ket{g}$. Therefore, we can utilize
also the two states $\ket{g}$ and $\ket{a}$ in order to encode a
qubit, which moreover, interacts with the cavity if there is one
photon in the cavity mode. Following Eq.~(\ref{e2b}), therefore, an
atom prepared in a superposition of $\ket{g}$ and $\ket{a}$ is
transformed due to
\begin{subequations}\label{transf}
\begin{eqnarray}
&& \ket{a,\bar{0}} \rightarrow \ket{a,\bar{0}}, \quad
   \ket{a,\bar{1}} \rightarrow \ket{a,\bar{1}}, \label{transf1} \\
&& \ket{g,\bar{0}} \rightarrow \ket{g,\bar{0}}, \quad
   \ket{g,\bar{1}} \rightarrow - \ket{g,\bar{1}} \label{transf2}
\end{eqnarray}
\end{subequations}
for the case of a full Rabi rotation $\Omega t = 2 \pi$ of the
atom-cavity state. Apparently, this transformation is the same as
the controlled-z gate (\ref{gates}) if the states $\{
\ket{a,\bar{0}}, \ket{a,\bar{1}}, \ket{g,\bar{0}},\ket{g,\bar{1}}
\}$ are taken as the basis. For this reason, we can use one full $2
\pi$ Rabi rotation to implement the single $A-M_2$ controlled-z gate
by carrying out the following three steps: If the atom is initially
prepared in a superposition of the $\ket{e}$ and $\ket{g}$ states,
we expose it to the two resonant $\pi$ Ramsey pulses, with the first
pulse being tuned to the $g \leftrightarrow a$ transition frequency
and the second pulse to $e \leftrightarrow g$. These two steps
transfer coherently the state of the qubit
\begin{equation}\label{transfer}
\alpha \ket{e} + \beta \ket{g} \;\rightarrow\; \alpha \ket{g} +
\beta \ket{a}, \qquad |\alpha|^2 + |\beta|^2 = 1.
\end{equation}
from the $\{\ket{e},\, \ket{g}\}$ into the $\{\ket{g},\, \ket{a}\}$
basis. After the transfer (\ref{transfer}) has been made, a $2\pi$
Rabi pulse is applied to the $A-M_2$ atom-cavity system that leads
to the transformations (\ref{transf}), or equivalently, to a
controlled-z gate between cavity mode $M_2$ and atomic qubit. Let us
note here that the cavity by Haroche and coworkers has a small hole
in the center of the upper cavity mirror which enables one to couple
a microwave source $S^\prime$ to an atom that moves through the
cavity [see Fig.~\ref{fig:1}(a)]. This microwave source, therefore,
can be used to act successively on both, the $g \leftrightarrow a$
and $e \leftrightarrow g$ atomic transitions and implement the
coherent transfer (\ref{transfer}).

The last operation we should discuss here, is the controlled-z gate
that acts upon the $M_1-M_2$ system and which is displayed in
Fig.~\ref{fig:4}(d) in the terms of ancilla qubits. This gate acts
on the cavity modes prepared in the state $\ket{+, \bar{+}}$
producing the entangled state
\begin{equation}\label{state0}
\frac{1}{2} \left[ \ket{(0+1),\bar{0}} + \ket{(0-1),\bar{1}}
\right].
\end{equation}
In fact, this state can be alternatively generated from the
initially empty cavity $\ket{0, \bar{0}}$ by means of one auxiliary
atom $A_s$ initialized in the excited state, which crosses the
cavity before the main chain of atoms. This is achieved if the
auxiliary atom first interacts for a $\pi/2$ Rabi pulse with the
mode $M_1$, then by a $\pi/2$ Ramsey pulse with the microwave source
$S^\prime$, and finally for a $\pi$ Rabi pulse with the mode $M_2$.
Using the Eqs.~(\ref{rot1}), (\ref{e1}) and (\ref{e2}), this
sequence of pulses produces the state
\begin{equation}\label{state1}
\frac{1}{2} \left[ \ket{(0+1),\bar{0}} + \im \ket{(0-1),\bar{1}}
\right] \, ,
\end{equation}
while the auxiliary atom $A_s$ is factorized out in its ground
state. In contrast to Eq.~(\ref{state0}), in the Eq.~(\ref{state1})
an extra factor $\im$ occurs because of the orthogonal polarization
of mode $M_2$ with respect to $M_1$, which however, is compensated
by the $A_{2N} - M_2$ mapping operation ($\pi$ Rabi pulse) as we
shall see below.

With this analysis of the individual (gate) operations, we have
established all ingredients that are needed in order to generate the
$2 \times N$ cluster state, and which are entirely adapted to our
cavity set-up. The overall scheme is displayed in
Fig.~\ref{fig:5}(c) in which the gates inside of the dashed box must
be repeated $N-3$ times. In addition to the notation we have used
before, the letter $R_c$ in this figure denotes the Ramsey zone
inside of cavity and associated to the microwave source $S^\prime$,
while the white circle refers to a Ramsey pulse that is tuned to the
atomic $g \leftrightarrow a$ transition frequency. Note that the
last $A_{2N} - M_2$ gate ($\pi$ Rabi pulse) maps the cavity states
$\ket{\bar{0}}$ and $\ket{\bar{1}}$ upon the atomic $\ket{g_{2N}}$
and $\ket{e_{2N}}$ states together with factorization of the cavity
mode $M_2$ in the state $\ket{\bar{1}}$. According to the
Eq.~(\ref{e2a}), moreover, such a mapping implies one extra factor
$\im$ if the cavity mode $M_2$ was empty, which together with the
$\im$ factor from Eq.~(\ref{state1}), gives rise to an irrelevant
global phase.

\begin{figure}
\begin{center}
\includegraphics[width=0.46\textwidth]{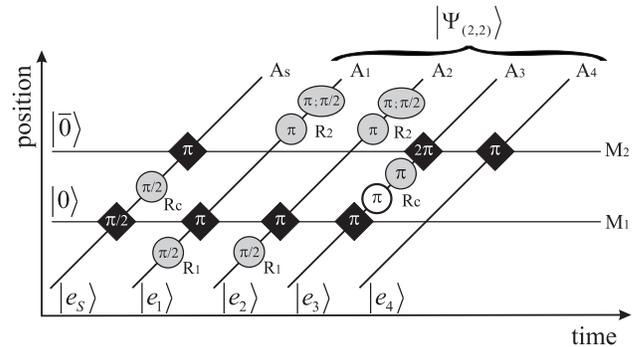}
\caption{Temporal sequence for the generation of the box state
(\ref{box}) that corresponds to the circuit from Fig.~\ref{fig:5}(c)
if $N=2$.}
\label{fig:6}
\end{center}
\end{figure}

To understand better the scheme for the generation of $2\times N$
cluster state, Fig.~\ref{fig:6} displays (the temporal sequence of)
all steps that are needed to generate the $2\times2$ box state
\begin{eqnarray}\label{box}
\ket{\Psi_{(2,2)}} = & \frac{1}{2} & \left( \ket{g_1, +_2, a_3, +_4}
                         + \ket{g_1, -_2, g_3, -_4} \right. \\[0.1cm]
                     &&  \left. + \ket{e_1, -_2, a_3, -_4} +
                         \ket{e_1, +_2, g_3, +_4}  \right) \notag
\end{eqnarray}
from Fig.~\ref{fig:4}(b). For the sake of brevity, neither the state
$\ket{g_s, 1, \bar{1}}$ of the auxiliary atoms and the cavity is
shown in this expression since they are both factorized out after
the sequence of steps has been completed. Obviously, the state
(\ref{box}) is equivalent to the state (\ref{box-def}) by making the
assignments
\begin{eqnarray}
\begin{array}{llll}
   \ket{g_1} = \ket{0_1}, & \ket{g_2} = \ket{0_2}, &
   \ket{a_3} = \ket{0_3}, & \ket{g_4} = \ket{0_4}, \\[0.1cm]
   \ket{e_1} = \ket{1_1}, & \ket{e_2} = \ket{1_2}, &
   \ket{g_3} = \ket{1_3}, & \ket{e_4} = \ket{1_4}.
   \end{array}  \notag
\end{eqnarray}
In this subsection, we have shown that each atom from a chain of
$2N$ uncorrelated atoms is incorporated into the $2 \times N$
cluster state by performing a Rabi $\pi$ (or $\pi$ followed by
$2\pi$) rotation and (if required) Ramsey pulses applied before,
inside, and/or behind the cavity.

\begin{figure}
\begin{center}
\includegraphics[width=0.46\textwidth]{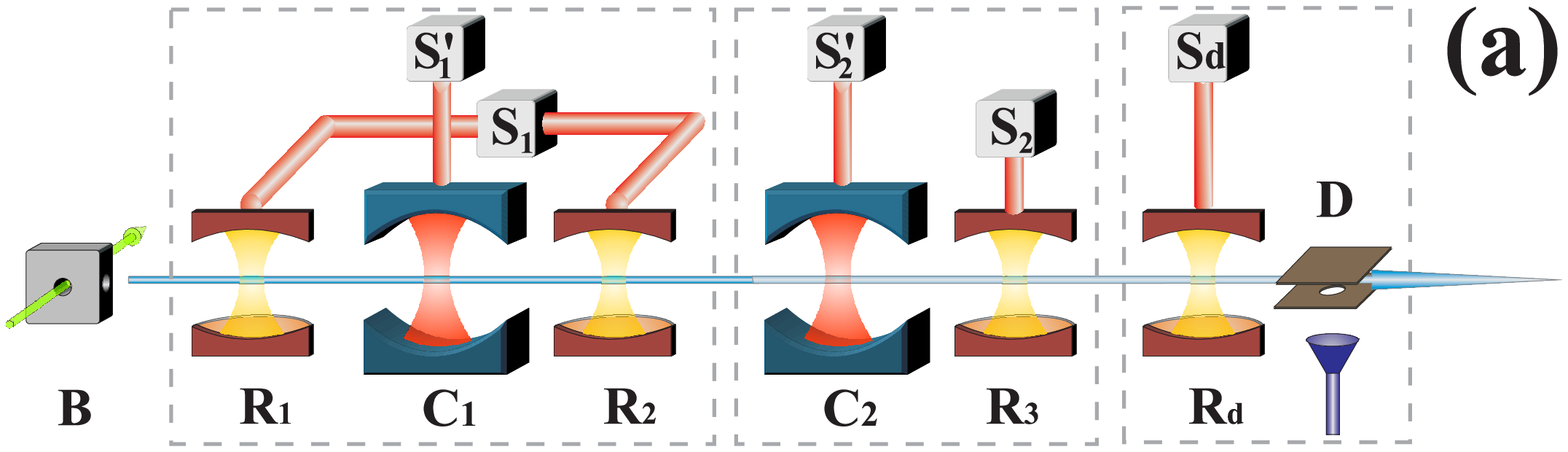} \\
\vspace{0.25cm}
\includegraphics[width=0.46\textwidth]{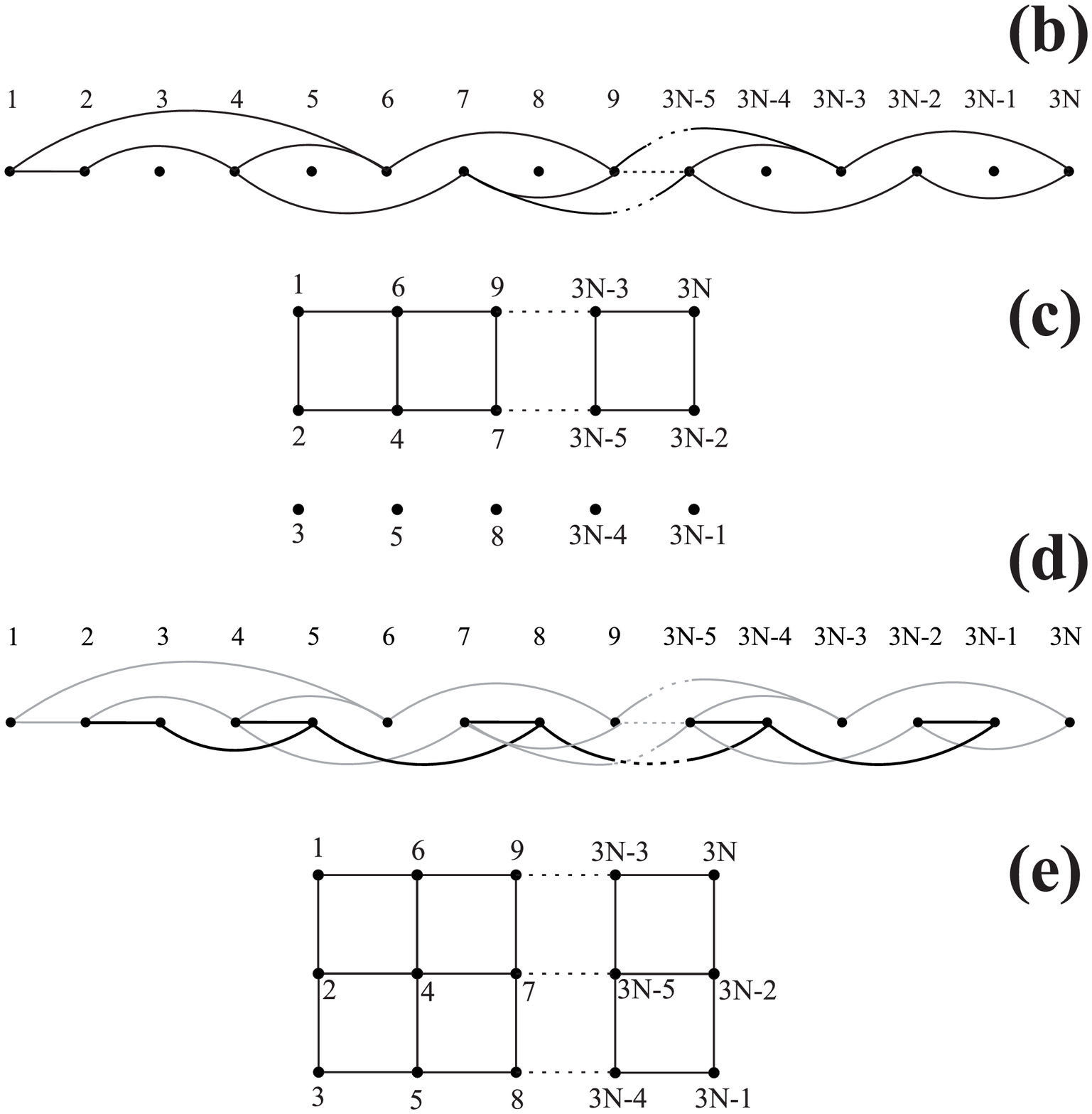}
\caption{(Color online) (a) Schematic set-up of an experiment with
two bimodal cavities $C_1$ and $C_2$. The classical fields in the
Ramsey zones before and after the cavities as well as those inside
of the cavities are generated by the microwave sources $S_1$,
$S^\prime_1$, $S_2$ and $S^\prime_2$. (b) Procedure for defining the
edges for a chain of $3N$ atoms (nodes) such that an effective
two-dimensional $2 \times N$ cluster state (c) is generated, where
the nodes from the third row are disconnected from all other atoms
of the chain. (d) Definition of the remaining edges which transform
the ($2 \times N$ cluster) state of the atomic chain into a $3
\times N$ cluster state (e) by making use of the second bimodal
cavity in the set-up.}
\label{fig:7}
\end{center}
\end{figure}

\subsection{$3 \times N$ Cluster State}

In the last two subsections, we have seen how to generate the $1
\times N$ cluster state by means of a single-mode cavity and how to
get a $2 \times N$ cluster state by using a bimodal cavity. Since
light has only two different polarization states, obviously, one
cannot use one similar technique to generate the $3 \times N$
cluster state by just utilizing a single cavity device.

In this subsection, instead, we shall present and explain a scheme
that enables one to generate a $3 \times N$ cluster state by using
an array of two bimodal cavities, i.e.\ simply by placing one
additional cavity $C_2$ and one Ramsey zone $R_3$ behind the zone
$R_2$, and before the detection area (at the Ramsey zone $R_d$) as
shown in Fig.~\ref{fig:7}(a). This scheme can be divided into the
following two steps: (i) implementation of the controlled-z gates
(edges) within a chain of $3N$ atoms according to
Fig.~\ref{fig:7}(b), which lead to the generation of a $2 \times N$
cluster state, such that the third row of atoms remains disconnected
from all other atoms as displayed in Fig.~\ref{fig:7}(c). Only the
cavity $C_1$ and microwave sources $S_1$ and $S^\prime_1$ are
utilized in this step, and the circuit that generates these edges is
the same as shown in Fig.~\ref{fig:5}(c) up to a re-assignment of
the atomic labels. All the atoms that remain disconnected during
this step simply pass through the first cavity being detuned from
the resonance ($\Delta = \delta$) with both cavity modes, and
therefore, without interacting with the cavity modes. (ii) The
second cavity $C_2$ and microwave sources $S_2$ and $S^\prime_2$ are
then utilized in order to create the additional (black) edges
according to Fig.~\ref{fig:7}(d). This step completes the generation
of the $3 \times N$ cluster state in which all neighbors are
connected to each other as displayed in Fig.~\ref{fig:7}(e).

\begin{figure}
\begin{center}
\includegraphics[width=0.46\textwidth]{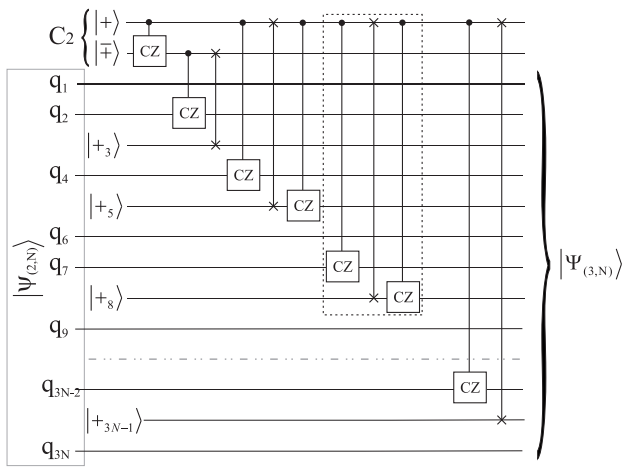} \\
\caption{Quantum circuit that performs the set of controlled-z gates
(edges) from Fig.~\ref{fig:7}(c), i.e.\ which transforms the
$\ket{\Psi_{(2,N)}}$ cluster state and the $N$ uncorrelated qubits
into the cluster state $\ket{\Psi_{(3,N)}}$ . The
$\ket{\Psi_{(2,N)}}$ state is given by qubits labeled as $q_1, q_2,
\ldots, q_{3N-2}, q_{3N}$ and the $N$ uncorrelated qubits are
labeled as $3, 5, \ldots, 3N-3, 3N-1$, respectively (see text for
details).}
\label{fig:8}
\end{center}
\end{figure}

Neither of these two steps do require any additional atom-cavity
gates that has not been described and discussed in the previous
subsections. Therefore, this procedure enables us to construct the
$3\times N$ cluster states in a way which is well adapted to our
set-up with two bimodal cavities. Because the first step can be
realized by applying the circuit from Fig.~\ref{fig:4}(d), we need
to display and discuss here only the circuit for the second step.
This is shown in Fig.~\ref{fig:8}, where the gates inside of the
dashed box must be repeated $N-4$ times. Similar as above, this
circuit can be translated in a straightforward way into a temporal
sequence of the atom-cavity interactions and single atomic gates.

By having understood the construction of the $3 \times N$ cluster
states, we can use the recipe from Fig.~\ref{fig:7} to generate the
two-dimensional regular cluster states of arbitrary size. Similarly
as the $3 \times N$ cluster state is obtained from the $2 \times N$
cluster and $N$ disconnected qubits, one can insert more cavities
into the experimental set-up in order to generate $M \times N$
cluster state (by means of $M-1$ cavities in total), and with a
proper assignment of the atomic labels to the nodes of the cluster
state.

Of course, there may arise the question of how many atoms from one
atomic chain can be incorporated in a cluster state, in line with
the recent developments in cavity QED? To obtain some rough
estimate, let us consider a scenario in which each atomic qubit can
be built into the cluster state for the price of a $\sim 3 \pi$ Rabi
rotation. If we assume that the (minimum) distance between any two
subsequent atoms in the chain is equal to the double waist length of
the cavity mode, then, the number of atoms is approximately related
to the lifetime $T$ of the atom-cavity system via relation
\begin{equation}\label{relation}
N \simeq \frac{1}{6} \frac{T}{T_{\pi}} \, \varepsilon \, ,
\end{equation}
where $T_{\pi}$ denotes the required time for a single $\pi$ Rabi
rotation and $\varepsilon$ is a factor which accounts for all
corrections to our idealized scheme. Such corrections may concern
the imperfect realization of the Rabi and Ramsey pulses, the
overlapping interaction of two atoms from the chain with the same
cavity mode, the effects of noisy channels and stray fields, and
others. In practise, such additional disturbances may lead to a much
smaller number $N$ of atoms that can be treated coherently. For the
atomic velocity $\upsilon = 500 \, m/s$, that was utilized in the
microwave cavity experiments by Haroche and coworkers, a single
$\pi$ Rabi rotation takes about $T_{\pi} \approx 10 \, \mu s$.
Moreover, the lifetime of the atom-cavity system is limited mainly
by the radiative lifetime of the atoms $T \simeq 30 \, ms$ (while
the cavity coherence time $\simeq 130 \, ms$ is much longer).
Therefore, by making a conservative estimate for the correction
factor in Eq.~(\ref{relation}), say $\varepsilon = 0.2$, we still
obtain $N \simeq 100$ atoms which may pass the cavity within the
given lifetime.

\subsection{Remarks on the Realization of Proposed Schemes}

Obviously, the realization of large (entangled) cluster states with
a trustworthy fidelity is an experimental challenge by itself. In
the earlier cavity-QED experiments by Rauschenbeutel and coworkers
\cite{sc288}, the generation of a three-partite
Greenberger-Horne-Zeilinger state (that is equivalent to the $N=3$
linear cluster state) was reported with a fidelity of $0.54$ \%{},
that is just above the threshold which is necessary to prove the
generation of this state. In this subsection, we shall therefore
discuss the main limitations that may arise experimentally in the
generation of larger cluster states by using microwave cavities
similar to those as applied in the Laboratoire Kastler Brossel (ENS)
in Paris.

One of the main bottleneck in cavity-QED experiments is the low
surface quality of the cavity mirrors, i.e.\ the local roughness and
the deviations from the spherical geometry. These defects cause the
scattering of photons outside the cavity mode and thus reduce the
coherent storage time of photons within the cavity. The storage time
of photons, in turn, limits the number of quantum operations (gates)
that can be performed successively before the atom-cavity state
becomes destroyed. This rapid loss of coherence during the evolution
of the atom-cavity state has stimulated Haroche and coworkers to
develop a new ultrahigh-finesse cavity devices \cite{apl90} for
which the quality factor was increased by about two orders of
magnitude. Such a high quality factor enables to perform more than
hundred quantum logical operations within the lifetime of a photon
inside the cavity (see estimations in the previous subsection).
Moreover, in order to minimize the contribution of thermal photons
that occur in microwave cavities due to thermal fields, the cavity
was cooled down to $0.8$ K which yields an average number of
$n_{\rm\, th} \simeq 0.05$ thermal photons. This ensures that the
contributions of thermal photons can be neglected for the evolution
of the cavity states.

In practice, the atoms that are emitted from the atomic source have
a spread in their velocities. For a given chain of atoms, such a
spread will lead to small deviations in the time intervals of the
atom-cavity interactions and will introduce uncertainties in the
duration of Rabi and Ramsey pulses. In order to minimize this
velocity spread, a velocity selector has been placed right after the
atomic source $B$ in experiments by Haroche and co-workers. This
selector reduces the velocity spread to $\sim 2$ m/s \cite{haroche1}
which being compared to the typical velocity of $500$ m/s of the
atoms, implies that the error due to the velocity spread is less
than one percent and negligible for most purposes. The small spread
in the velocities gives rise also to a small spatial dispersion
($\lesssim 1$ mm) of the atomic positions while they pass through
the cavity. This spatial dispersion compared to the resonant cavity
wavelength ($\sim 5.9$ mm) implies that only a small deflection from
the cavity antinode may occur and may yield a similar small
deviation of the atom-cavity coupling from its nominal value.

Indeed, the control and manipulation of the resonant cavity
frequency is essential in order to achieve a resonant atom-cavity
interaction regime. Any deviation from this resonant regime would
lead to spurious matrix elements in the atom-cavity gates
(\ref{m-swap}) and (\ref{i-swap}). In the experiments by Haroche and
co-workers, the cavity frequency is manipulated by slightly changing
the distance between the cavity mirrors. By making use of a
piezoelectric stack placed under the lower cavity mirror, a fine
tuning of the cavity length was achieved with $\sim 1$ MHz range,
which has to be compared to the atomic transition frequency $51.099$
GHz. Again, such small deviations in the length seems to be
negligible. In the present work, however, we consider a bimodal
cavity scenario in which the atomic transition frequency is tuned to
the first or second cavity mode by applying a time-varying electric
field across the cavity gap, such that the required (Stark) shift of
the atomic $e \leftrightarrow g$ transition frequency is obtained.
As explained in the beginning of this section, a rather smooth
switch of the atom-cavity detuning is produced within the finite
time of $\sim 1 \, \mu s$ that could affect the evolution of the
cavity states \cite{jpb} whenever the switching pulse is comparable
to the Rabi pulse.

Finally, in order to describe the real evolution of the atom-cavity
system, one would have to include also the interaction with the
environment that has been omitted from the present considerations.
As pointed out in the beginning of this section, the resonant
atom-cavity interaction regime implies a negligible dissipation of
the cavity field during all the atom-cavity interaction time
periods. By setting sufficient distance between the atoms in a
chain, events with several atoms in the same cavity mode are
avoided. However, such a distance may yield the cavity to be empty
and thus field dissipation would have to be taken into account. A
detailed investigation of the evolution of bimodal cavity states
with dissipation has been performed by Magalhaes and Nemes
\cite{pra70} and the dynamics of the cavity-field relaxation has
been understood. On the other hand, a quantum simulator has been
developed in our group \cite{cpc175} which could be utilized in
future studies to investigate the role of decoherence on the
evolution of cluster states. Overall, we conclude that further
improvement of the cavity mirrors quality factor and the
time-varying electric field characteristics are needed in order to
produce large cluster states by the suggested schemes with a good
fidelity.

\section{Summary and Outlook}

In this work, two novel schemes are presented to generate the
two-dimensional $2 \times N$ and $3 \times N$ cluster states within
the framework of cavity QED. These schemes are based on the resonant
interaction of a chain of Rydberg atoms with one or two bimodal
cavities, i.e.\ cavities that support two independent modes (with
orthogonal polarization) of the photon field. In addition, we have
shown and discussed how the scheme for the $3 \times N$ cluster
state can be extended also for the construction of two-dimensional
$M \times N$ cluster states of arbitrary size. This is achieved by
using $M-1$ bimodal cavities in a row. Using the graphical language
of temporal sequences and quantum circuits, a comprehensive
description of all necessary gates and manipulations is provided. We
stress that the given scheme(s) can easily be adapted to the
present-day microwave cavity QED experiments although their
realization is still a challenge, especially if one is interested in
cluster states with $N > 4$.

The results of this paper also suggest that cavity QED provides a
suitable framework not only for the generation of cluster states but
also for one-way quantum computations which are performed by a
sequence of single-qubit projective measurements  (with possible
feedforwarding). For these computations, a cluster state of
appropriate size is needed together with two types of measurements
\cite{BR2}: (i) measurement in the basis $\{ \ket{0_k}; \ket{1_k}
\}$ and (ii) measurement in the basis
\begin{equation}\label{basis}
\{ \left(\ket{0_k} + e^{\im \varphi} \ket{1_k} \right) /\sqrt{2}; \,
\left(\ket{0_k} - e^{\im \varphi} \ket{1_k} \right) /\sqrt{2} \} \,
,
\end{equation}
where $\varphi$ denotes a real number. Those qubits which are not
projected (measured) finally, encode the output quantum state.

To see how the one-way quantum computations fit into the framework
of cavity QED, we can reconsider the set-up displayed in
Fig.~\ref{fig:1}(a). In this figure, a particular cluster state is
generated within a chain of atoms emitted by the atomic source $B$
after it crosses the cavity and microwave sources. This cluster
state, being encoded in the atomic chain, then enters into the
detection region, where each (Rydberg) atom is projected upon one of
its levels $e$, $g$, or $a$ and by which, therefore, the measurement
in the basis (i) is performed. In order to perform also the
measurement in the basis (ii), a Ramsey zone $R_d$ is installed and
applied before the detector. It can be shown \cite{prl79, sc288}
that a $\pi/2$ Ramsey pulse detuned by the value of $\varphi$ from
atomic resonance followed by a detection of the atom in the basis
(i), is equivalent to a projective measurement in the basis
(\ref{basis}). The last (unmeasured) atoms inside the atomic chain,
encode the final output quantum state. We may conclude, therefore,
that all necessary ingredients are available in order to perform
one-way quantum computations in the framework of cavity QED,
including also the preparation of the cluster state in the same
set-up.

\begin{acknowledgments}
This work was supported by the DFG under the project No. FR 1251/13.
\end{acknowledgments}

\end{document}